\documentclass[%
 reprint,
 amsmath,amssymb,
 aps,
]{revtex4-2}

\usepackage{graphicx}% Include figure files
\usepackage{xcolor}
\usepackage{dcolumn}% Align table columns on decimal point
\usepackage{bm}% bold math
\begin{document}

\preprint{APS/123-QED}

\title{Multi-parameter study of two-axis Hanle magnetometry on the Cs D$_1$ line}% Force line breaks with \\
%\title{Manuscript Title:\\with Forced Linebreak}% Force line breaks with \\

\author{A. Mozers}
 \email{arturs.mozers@lu.lv}
\author{A. Nikolajevs}
\author{F. Gahbauer}
\author{M. Auzinsh}
\affiliation{
 Laser Centre, University of Latvia, Rainis Boulevard 19, LV-1586 Riga, Latvia
}

\date{\today}% It is always \today, today, but any date may be explicitly specified

\begin{abstract}
We have studied experimentally and theoretically the impact of various parameters on the characteristics of signals from a two-axis Hanle magnetometer based on the Cs D$_1$ transition illuminated by linearly polarized pump and probe beams propagating through the cell in such a way that the probe polarization vector can be switched between the $zy$- and the $xz$-plane in order to measure the magnetic field along the $x$- and $y$-axes, respectively. When the probe polarization additionally is reflected about the pump polarization vector and the two resulting absorption signals are subtracted, dispersion signals centered around zero are obtained. These dispersion signals can be reproduced accurately by the theoretical model, and the information from experiment and modeling used to optimize the magnetometer characteristics.

\end{abstract}

\maketitle

%%%%%%%%%%%%%%%%%%%%%%%%%%%%%%%%%%%%%%%%%%%%%%%%%%%%%%%%%%%%%%%%%%%%%%%%%%%%%%%%%%%%%
\clearpage
\section{\label{sec:level1}Introduction\protect}
A two-axis Hanle magnetometer whose configuration permits using an optical cell with a single window has recently been proposed using $^4$He\cite{LeGal:2019}. 
The Hanle effect can be observed in a system that experiences hyperfine splitting.  Resonant optical pumping creates coherences among the magnetic sublevels, which are degenerate in the absence of external fields. These coherences are destroyed as an external magnetic field lifts the sublevel degeneracy~\cite{Hanle:1924,MoruzziStrumia,Alnis:2003}.
Resonances associated with this effect are significantly narrower in the ground state, where the relaxation times are longer, and it was first observed in 1964 using circularly polarized light~\cite{Lehmann:1964} and later using linearly polarized light (see ~\cite{Budker:2002} for a review).
In a Hanle magnetometer, one observes the absorption or fluorescence of a weaker probe beam that passes through the polarized atomic ensemble created by the pump beam~\cite{Laloe:1969, Gilles:2001}, yielding very sensitive atomic magnetometers~\cite{Happer:1977,Allred:2002,Budker:2007,Papoyan:2017}.
The typical approach is to create an orientation of atoms using circularly polarized light.
The polarized angular momentum distribution created in this way begins to precess about the external magnetic field, and the resulting change in polarization affects the absorption of a linearly polarized probe beam that propagates perpendicularly to the pump. 
However, instead of creating orientation with circularly polarized light, it is possible also to create alignment with linearly polarized light~\cite{Shi:2018a,Shi:2018b}.
The aligned angular momentum distribution will also precess about an external magnetic field perpendicular to it, which allows magnetometry as well~\cite{Breschi:2012}. 

\begin{figure}[htb!]
    \centering
    \includegraphics[width=0.6\linewidth]{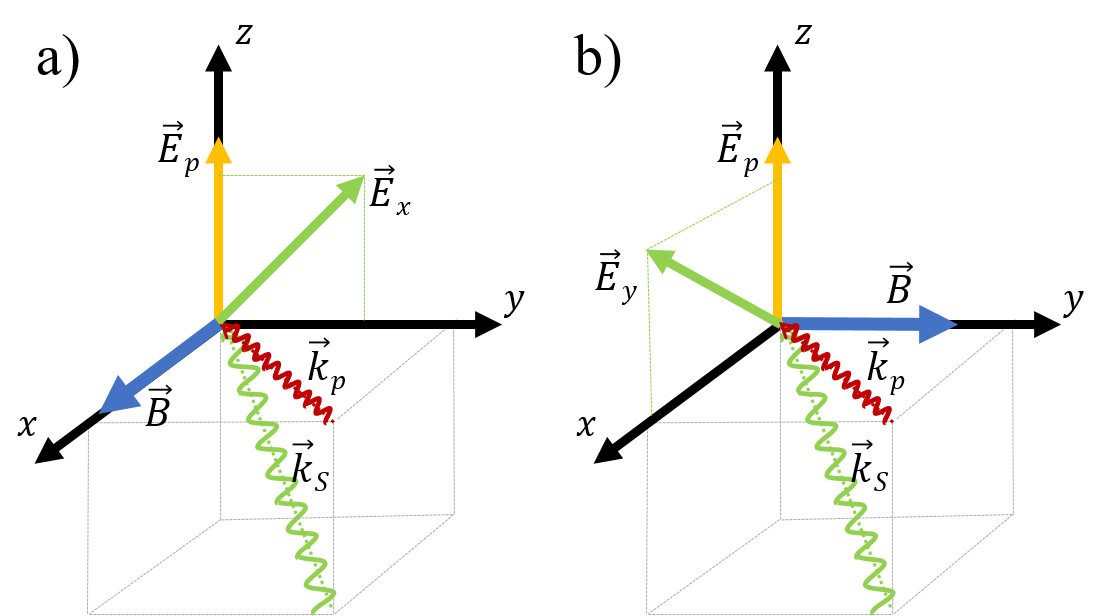}
    \caption{Excitation geometry. The pump beam polarization and propagation vectors are denoted by the subscript $p$. The probe beam propagation is denoted by the subscript S. The probe beam polarization vectors are denoted by the subscripts (a) $x$ and (b) $y$, for measuring the magnetic field $B$ along the $x$ or $y$ directions, respectively.}
    \label{fig:geometry}
\end{figure}

The advantage of the alignment Hanle magnetometer is that the pump and probe beams no longer must be perpendicular, but can be separated by a smaller angle. LeGal \textit{et al.} exploited this fact to propose a two-axis magnetometer, which required optical access only along a single axis~\cite{LeGal:2019,LeGal:2021,LeGal:2022}.
The basic geometry is shown in Fig.~\ref{fig:geometry}. With the linear polarization of the pump beam constant along the $z$-axis, the probe beam can reveal the the magnitude of the magnetic field along the $x$-axis when the linear polarization of the probe is in the $yz$-plane, and so its absorption depends on the rotation rate of the alignment induced by the pump beam around the $x$-axis. Note that in this paper the quantization axis is assumed to be always in the direction of the magnetic field.
The same applies \textit{mutatis mutandi} when the linear polarization of the probe beam is rotated onto the $xz$-plane. This change in the rotation of the probe polarization can be accomplished, by an electro-optic modulator (EOM).

In this work we studied the configuration proposed by LeGal \textit{et al.}~\cite{LeGal:2019} for Helium in the context of the Cesium D$_1$ line with a view to obtaining the optimum parameters for this medium. As a result, we studied the impact of fundamental magnetometer parameters, such as the specific transition, the pump and probe laser power, and ground-state relaxation on the signals which will affect magnetometer performance, such as signal amplitude and width. 
The paper is structured as follows. First, we describe the experimental setup of a Cesium D$_1$ Hanle magnetometer together with some experimental and data processing techniques that helped to improve the signals.
Then we present a theoretical model that was used to calculate absorption and fluorescence signals of a magnetometer based on a pump and a probe beam.
The model also allowed calculating the probability distributions of atomic angular momentum.
Finally, we present results of our multi-parameter study and make recommendations for a future Hanle magnetomer. 

%%%%%%%%%%%%%%%%%%%%%%%%%%%%%%%%%%%%%%%%%%%%%%%%%%%%%%%%%%%%%%%%%%%%%%%%%%%%%%%%%%%%%
\clearpage
\section{\label{sec:experiment}Experiment}
\begin{figure}[htb!]
    \centering
    \includegraphics[width=\linewidth]{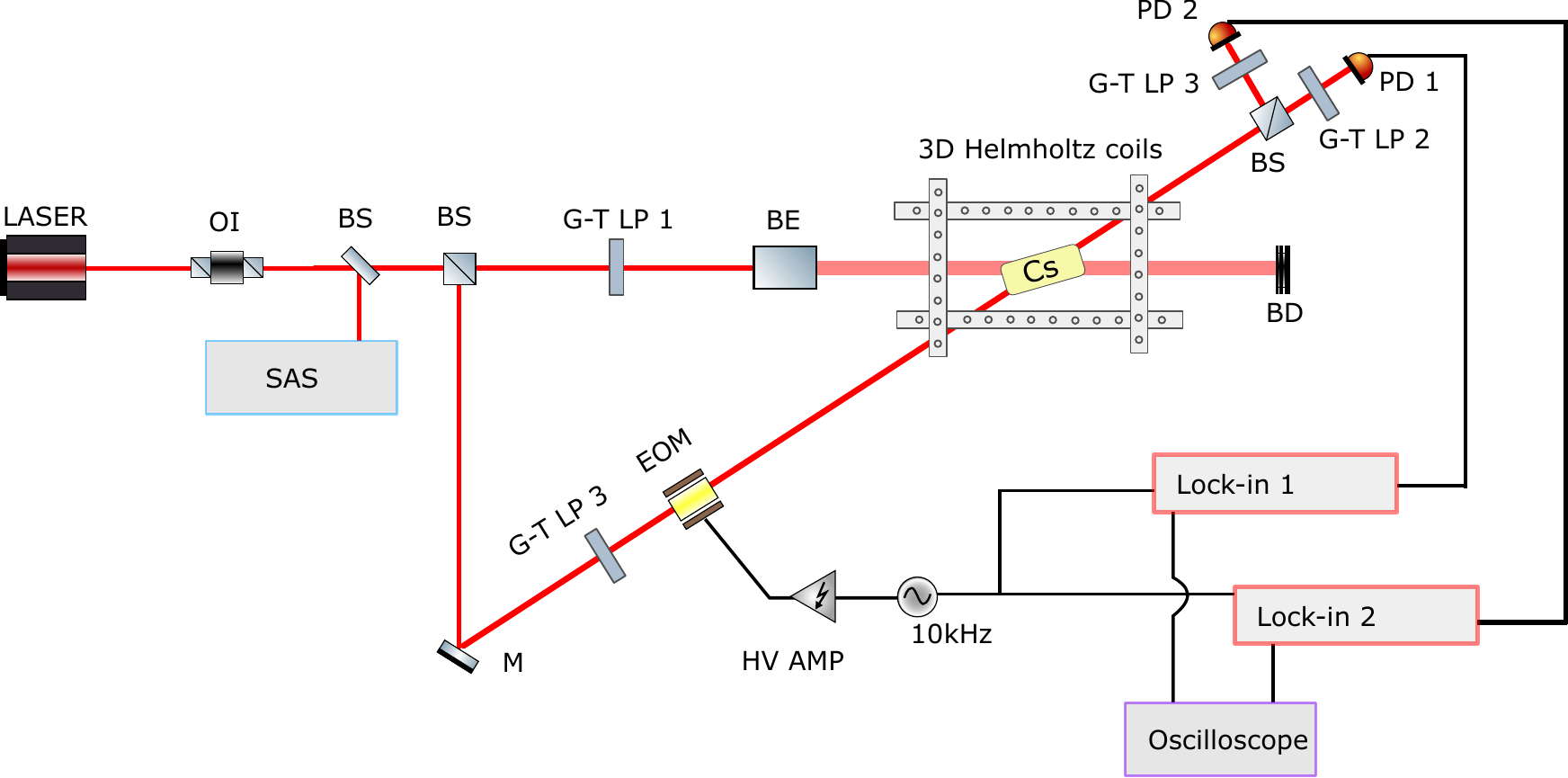}
    \caption{Experimental setup. Cs---Cesium cell; OI---optical isolator; BS---beam splitter; SAS---seturation absorption setup; G-T LP---Glan-Thompson polarizer; BE---Beam expander; BD---beam dump; EOM---electro-optical modulator; PD---photodiode}
    \label{fig:experiment}
\end{figure}

This geometrical arrangement (see Fig.~\ref{fig:geometry}) of magnetic field and light polarization vectors is implemented by the experimental setup shown schematically in Fig.~\ref{fig:experiment}.
Pump and probe beams are produced by a single laser.
The laser frequency is stabilized on the peak of a saturated absorption spectrum.
A beam splitter divides the beam into pump and probe beams.
Both pass through Glan-Thompson polarizers to create linearly polarized beams.
The pump beam passes through a beam expander so that different beam sizes can be tested.
The probe beam passes through an electro-optical modulator (EOM) with a 30$^{\circ}$ angle between the polarization vector of the incoming light and the crystal axis, which allows the light polarization vector to be rotated by 60$^{\circ}$ if the phase retardance is half-wave. This procedure allows us to switch between the two geometries shown in figure~\ref{fig:geometry}.
The system is configured so that in one case, the polarization vector of the probe light is confined in the $yz$ plane and is sensitive to a static magnetic field along the $x$-axis (see Fig.~\ref{fig:geometry}(a).
In the other case, the polarization of the probe beam is located in the $xz$-plane and sensitive to the external magnetic field along the $y$-axis.
After the cell, the probe beam is split again with a non-polarizing beam splitter and directed onto two photodiodes. The polarizers placed before the phodiodes enable the separation of the to probe beam polarizations by rejecting the unwanted polarizations for each direction.
The phase retardance can be modulated, and the modulation frequency can be fed into two lock-in amplifiers as a reference.
Then, one lock-in amplifier measuring the in-phase component will output the Hanle signal for the magnetic field along the $x$-direction, whereas the other lock-in amplifier, measuring the out-of-phase component, will measure the Hanle signal for the magnetic field along the $y$-direction.
The signals take the form of dispersion-type signals (see~Fig.~\ref{fig:transition-dependence}).

%%%%%%%%%%%%%%%%%%%%%%%%%%%%%%%%%%%%%%%%%%%%%%%%%%%%%%%%%%%%%%%%%%%%%%%%%%%%%%%%%%%%%
\clearpage
\section{\label{theory}Theoretical Model}
The theoretical model we use to describe our results is based on a numerical solution of the Optical Bloch Equations~\cite{Blushs:2004} and has been described in detail elsewhere (see, for example, \cite{Auzinsh:2008}.)

We describe the state of the ensemble of atoms using the density operator $\rho$,
\begin{equation}
    \rho = \sum_j p_j \left| \psi_j \right> \left< \psi_j \right|,
\end{equation}
where the coefficients $p_j$ are the probabilities of finding the system in the pure states $\left| \psi_j \right>$.

Then we build a system Hamiltonian out of the Hamiltonian of the bare atom $\hat{H}_0$, the Hamiltonian for the interaction with the external magnetic field $\hat{H}_B$, and the Hamiltonian for the interaction with the electromagnetic field $\hat{V}$:
The Hamiltonian is given by
\begin{equation}
    \hat{H}=\hat{H_0}+\hat{H_B}+ \hat{V}.
\end{equation}
Using this Hamiltonian and adding an operator that describes relaxation $\hat{R}\rho$, the optical Bloch equations (OBEs) describe the evolution of the density matrix as follows:
\begin{equation}
    i\hbar \frac{\partial \rho}{\partial t} = [\hat{H},\rho] + i\hbar \hat{R}\rho.
\end{equation}

In order to make the Hamiltonian tractable, we apply the rotating wave approximation, and make a series of assumptions we can eliminate the optical coherences and simplify the OBEs~\cite{Blushs:2004}: 

\begin{align}
\label{eq:zc}
\frac{\partial \rho_{g_ig_j}}{\partial t} = \sum_{e_k,e_m}\left(\Xi_{g_ie_m} + \Xi_{g_je_k}^{\ast}\right)d_{g_ie_k}^\ast d_{e_mg_j}\rho_{e_ke_m}  \nonumber\\  
-\sum_{e_k,g_m}\Big(\Xi_{g_je_k}^{\ast}d_{g_ie_k}^\ast d_{e_kg_m}\rho_{g_mg_j}  \nonumber\\  
+\Xi_{g_ie_k}d_{g_me_k}^\ast d_{e_kg_j}\rho_{g_ig_m}\Big)  \nonumber\\ 
- i \omega_{g_ig_j}\rho_{g_ig_j} +  \sum_{e_ke_l}\Gamma_{g_ig_j}^{e_ke_l}\rho_{e_ke_l} \nonumber \\ 
- \gamma\rho_{g_ig_j}  + \lambda\delta(g_i,g_j) 
\\
\frac{\partial \rho_{e_ie_j}}{\partial t} = \sum_{g_k,g_m}\left(\Xi_{g_me_i}^\ast + \Xi_{g_ke_j}\right)d_{e_ig_k} d_{g_me_j}^\ast\rho_{g_kg_m} -\nonumber \\
-\sum_{g_k,e_m}\Big(\Xi_{g_ke_j}d_{e_ig_k} d_{g_ke_m}^\ast\rho_{e_me_j}  \nonumber \\ 
+\Xi_{g_ke_i}^\ast d_{e_mg_k} d_{g_ke_j}^\ast\rho_{e_ie_m}\Big)  + \nonumber \\ 
-i \omega_{e_ie_j}\rho_{e_ie_j}  -  (\Gamma + \gamma)\rho_{e_ie_j},
\end{align}

where 
\begin{equation}
    \Xi_{g_i e_j}=\frac{\Omega_R^2}{\frac{\Gamma+\gamma+\Delta \omega}{2}+ i\left(\overline{\omega}-\mathbf{k}_{\overline{\omega}}\cdot \mathbf{v}+ \omega_{g_i e_j} \right)}.
\end{equation}

In this equation, the Rabi frequency is given by
\begin{equation}
\Omega_R^2=\left( \frac{|\varepsilon_{\overline{\omega}}|}{\hbar} \right)^2 \left| \left< J_e || d || j_g \right> \right|^2.
\end{equation}
The dipole transition matrix elements between states $\left| j \right>$ and $\left< i \right|$ are given by $d_{ij}=\left< i | \mathbf{\hat{d}} \cdot \mathbf{e} | j \right>$, where $\mathbf{\hat{d}}$ is the dipole operator and $\mathbf{e}$ is the electric field vector of the light field. The energy splitting between ground-state sublevels are given by $\omega_{g_ig_j}$; those between excited-state sublevels by $\omega_{e_ie_j}$. 

The relaxation term is given by
\begin{equation}
    \Gamma^{e_k e_l}_{g_i g_j}=(2J_e + 1)\Gamma \sum_{q=-1}^{1} (-1)^q \left< e_k | d_q | g_i \right> \left< g_j | d_{-q} | e_l \right>,
\end{equation} 
The fact that coherences are carried by one photon is expressed by the equation  $m_{e_i}-m_{g_i}=m_{e_j}-m_{g_j}$. Fly-through relaxation, which takes into account atoms flying out of the laser beam, is represented by $\gamma$, and it is given by
\begin{equation}
    \gamma \propto \overline{v}/d_{beam},
    \label{eq:small_gamma}
\end{equation}
where $\overline{v}$ is the average thermal velocity of the atoms and $d_{beam}$ is the effective beam diameter. The term $\lambda$ represents the atoms that fly into the beam. The term $\Gamma$ represents the spontaneous relaxation of atoms from the excited state. 

The Doppler effect is taken into account by the term $\mathbf{k}_{\overline{\omega}}\cdot \mathbf{v}$.  The results of this calculation must be averaged over the Doppler velocity profile. 
Furthermore, the levels used in the above equations are mixed by external fields, which must be taken into account when calculating these terms. 

Once we have calculated the density matrix, we can use it to calculate the absorption (and fluorescence) of light of a given polarization vector.

The  absorption of a weak probe beam of polarization $\epsilon_\text{probe}$ is given by
\begin{equation}
    A(\mathbf{e}_\text{probe}) = \tilde{A}_0 \sum_{e_ig_jg_k} \frac{d_{e_ig_j}^\text{(probe)} \rho_{g_jg_k} d^{\ast\text{(probe)}}_{g_k e_i} }{\Delta_{e_ig_j}^2 + \left (\frac{\Gamma+\gamma+\Delta \omega}{2}\right )^2},
\end{equation}
where the dipole transition matrix for a photon with the polarization $\mathbf{e}_\text{probe}$ is given by $d_{e_kg_i}^\text{(probe)} = \langle {e_k} | {\bf\hat d} \cdot {\mathbf{e}_\text{probe}} | {g_i} \rangle$
The term $\tilde{A}_0$ is a constant of proportionality.

Knowing the density matrix also allows one to calculate the distribution of angular momentum of an atomic state and to visualize it in three dimensions, which yields further insight into what is going on. This probability is expressed as a surface in 3-dimensional space as follows~\cite{Auzinsh:2010book}:
\begin{equation}
    \rho_{FF}(\theta,\phi)=\sqrt{\frac{4\pi}{2J+1}}\sum_{\kappa=0}^{2F}\sum_{q=-\kappa}^{\kappa} \left< F F \kappa 0 | FF \right> \rho^{\kappa q}Y_{\kappa q}(\theta,\phi).
\end{equation}
In this equation, the polarization moments $\rho^{\kappa q}$ are 
\begin{eqnarray}
    \rho^{\kappa q} &=&\mathrm{Tr}\left(\rho \mathcal{T}_{q}^{\kappa}\right) \nonumber \\
                    &=&\sum_{mm'}\rho_{mm'}\left(\mathcal{T}_q^{\kappa} \right)_{mm'} \nonumber \\
                    &=&\sum_{mm'}(-1)^{F-m}\left< F m' F, -m| \kappa q\right>\rho_{mm'},
\end{eqnarray}
where the $\mathcal{T}_q^{\kappa}$ are irreducible tensor operators, or polarization operators.

%%%%%%%%%%%%%%%%%%%%%%%%%%%%%%%%%%%%%%%%%%%%%%%%%%%%%%%%%%%%%%%%%%%%%%%%%%%%%%%%%%%%%
\clearpage
\section{\label{results}Results \& Discussion}

\subsection{Dependence on Hyperfine transition Cs D$_1$}

\begin{figure}[htb!]
    \centering
    \includegraphics[width=\linewidth]{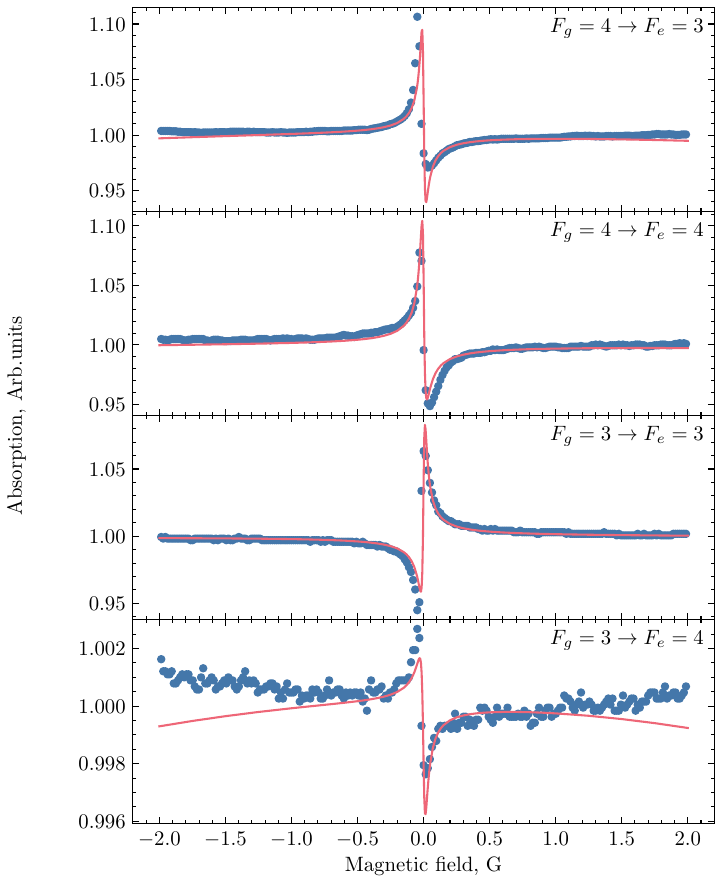}
    \caption{Points - Measured probe laser beam absorption versus magnetic field for the four transitions of the $^{133}$Cs $D_1$ line. Probe diameter $d_{probe}=2.186$~mm, pump diameter ${d_{pump}=21.425}$~mm. Solid curves - Calculated absorption versus magnetic field for the four transitions of the $^{133}$Cs $D_1$ line. Pump $\Omega_R=0.5$~MHz, $\gamma=0.007$~MHz. Note the different vertical scales.}
    \label{fig:transition-dependence}
\end{figure}

The first design parameter of any atomic magnetometer is the atomic transition that will be used, and hence experimental measurement (blue dots) of the probe absorption for all four transitions of the Cs D$_1$ line are shown in Fig.~\ref{fig:transition-dependence}.
The results of the model calculations are shown in Fig.~\ref{fig:transition-dependence} as solid red lines.
The probe transmission amplitude and width depends on the transition, and it is obvious that the $F_g=3\longrightarrow F_e=4$ transition would be the least suitable for magnetometry because of the small amplitude, which is not compensated by a particularly narrow width.

The strongest signal can be expected from the $F_g=4\longrightarrow F_e=3$ transition.
The signal from the $F_g=4\longrightarrow F_e=4$ transition shows a very similar signal amplitude, while the signal amplitude from the $F_g=3\longrightarrow F_e=3$ transition shows a reduction of approximately one third relative to the $F_g=4\longrightarrow F_e=3$ transition.

\begin{figure}[htb!]
    \centering
    \includegraphics[width=\linewidth]{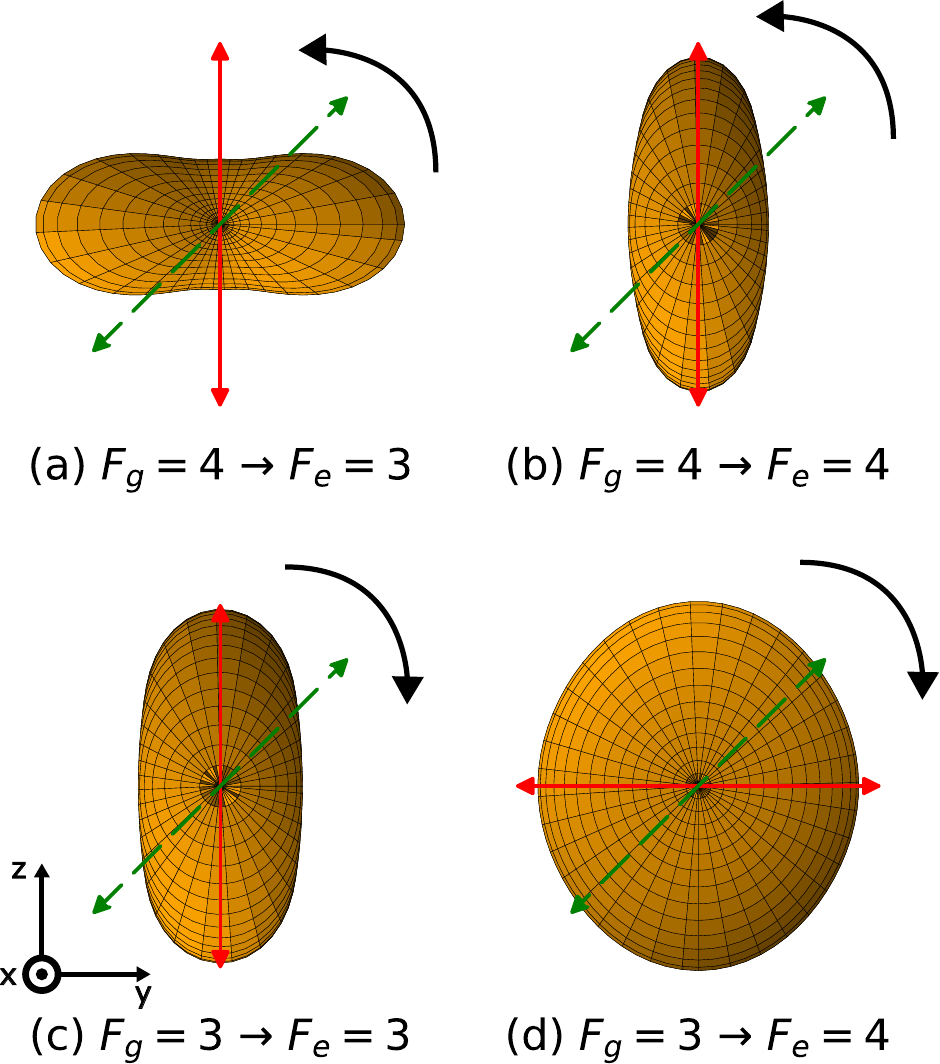}
    \caption{Surfaces of the ground-state angular momentum probability distributions for the different hyperfine transitions calculated with the same parameters as the results of Fig.~\ref{fig:transition-dependence}. The Doppler effect is not taken into account and atoms are presumed to be at rest with respect to the laser radiation. The red arrow shows the principal excitation direction of the transition dipole whereas the green arrow shows the polarization of the probe beam. The magnetic field is zero for these images, but the scan direction is perpendicular to the page.}
    \label{fig:AMPS}
\end{figure}

The difference in the signal shapes reflect differences in the underlying probability distributions of ground-state angular momentum, which are plotted in Fig.~\ref{fig:AMPS}.
The angular momentum probability distribution that corresponds to the $F_g=3\longrightarrow F_e=4$ transition is the most symmetric (see Fig.~\ref{fig:AMPS}(d)), and, as a result, leads to the signal with the lowest amplitude.
The angular momentum probability distribution for the $F_g=4\longrightarrow F_e=3$ transition (see Fig.~\ref{fig:AMPS}(a)) shows the greatest anisotropy of the four transitions, and so its signal amplitude and width will yield the greatest sensitivity to a magnetic field in a direction perpendicular to the page.

The different slope coefficients of the absorption signal at zero magnetic field can also be explained by the angular momentum probability distribution surfaces from figure~\ref{fig:AMPS}.
In figure~\ref{fig:transition-dependence} transitions $F_g=4\longrightarrow F_e=3$ and $F_g=4\longrightarrow F_e=4$ show negative slope coefficients, while it is positive for the $F_g=3\longrightarrow F_e=3$ transition.
This occurs due to the relative orientation of the principal excitation direction of the transition dipole (represented as a red arrow in fig.~\ref{fig:AMPS}) being either parallel or perpendicular with respect to the ground-state angular momentum distribution symmetry axis.
If the total angular momentum does not change $\Delta F=0$, the dipole moment of the transition is along the distribution of the angular momentum of the atoms (see fig.~\ref{fig:AMPS}(b) and (c)), but when $\Delta F=\pm1$ transitions are excited they are perpendicular (see fig.~\ref{fig:AMPS}(a) and (d)) (see~\cite{Auzinsh:2005MolPol} and references therein).
Additionally the two ground-state hyperfine levels of Cs $F_g=4$ and $F_g=3$ have opposite signs of the Land\'e factor.
Therefore, as the magnetic field is increased, the direction of the dipoles (red arrow in fig.~\ref{fig:AMPS}) of the excited atoms start to precess either towards or away from the polarization of the probe beam (green arrow in fig.~\ref{fig:AMPS}) depending on the sign of the Land\'e factor of the corresponding ground state hyperfine level. For negative magnetic field values, the precession occurs in the opposite direction.
Despite the fact that the dipole moments of all three transitions are aligned along the same axis (see fig.~\ref{fig:AMPS}(a), (b) and (c)), the slope of the $F_g=3\longrightarrow F_e=3$ transition has a positive slope in the signal due to the opposite precession of the dipole moment.
The dipoles of the $F_g=3\longrightarrow F_e=4$ transition start to precess in the same direction as dipoles of the $F_g=3\longrightarrow F_e=3$ transition (in the clockwise direction as represented in fig.\ref{fig:AMPS}), but because the initial direction of the dipole is perpendicular to the angular momentum distribution the dipoles precess away from the probe beam leading to a decrease in the absorption signal for positive magnetic field values.
As the magnetic field exceeds approximately 0.5 G the absorption signal reaches a constant value because the precession frequency of the dipoles is high enough for many revolutions of the angular momentum distribution to occur before it decays, which in essence means that the anisotropic angular momentum distribution has become isotropic in the plane perpendicular to the applied magnetic field direction.

Since the absorption signal amplitudes are very similar for both $F_g=4\longrightarrow F_e=3$ and $F_g=4\longrightarrow F_e=4$ transitions, as an example, in the subsequent subsections we provide analysis of the $F_g=4\longrightarrow F_e=4$ transition.

%%%%%%%%%%%%%%%%%%%%%%%%%%%%%%%%%%%%%%%%%%%%%%%%%%%%%%%%%%%%%%%%%%%%%%%%%%%%%%%%
\clearpage
\subsection{Subtracting the Probe beam} \label{sec:subtraction}
\begin{figure}[htb!]
    \centering
    \includegraphics[width=\linewidth]{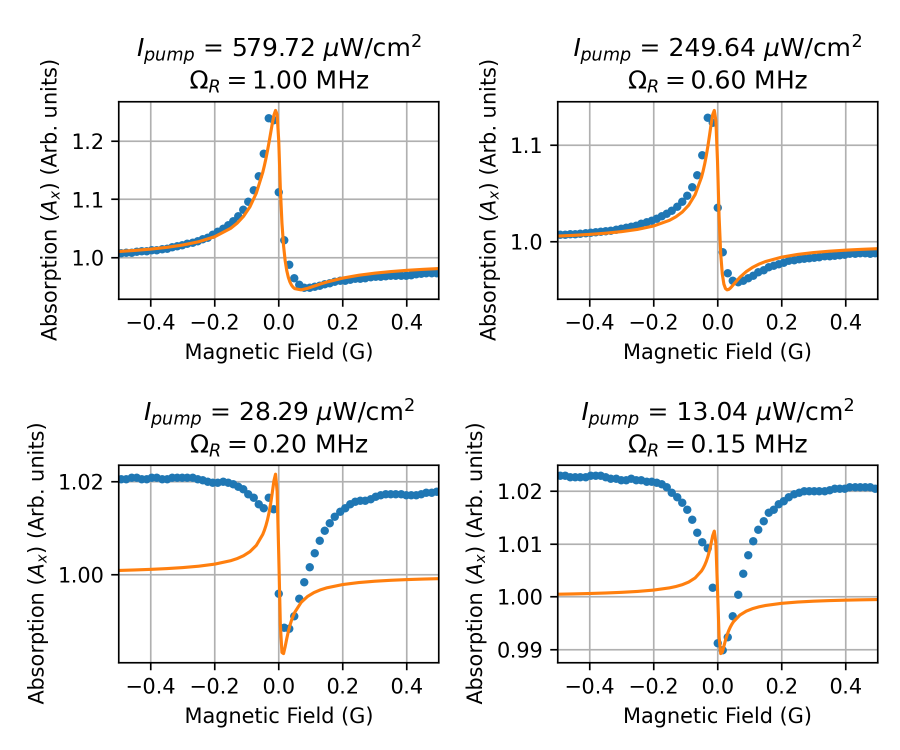}
    \caption{Absorption signals of individual probe polarization component ($A_x$) for various pump intensities for the $^{133}Cs~D_1$ $F_g = 4\rightarrow F_e = 4$ transition. Blue dots show the experimental measurements and the solid, orange line shows the results of experimental calculations. Note the different vertical scales for each subfigure.}
    \label{fig:no-subtraction}
\end{figure}

Figure~\ref{fig:no-subtraction} shows the probe beam absorption as a function of magnetic field for the $F_g=4\longrightarrow F_e=4$ transition of the Cs D$_1$ line for different values of the pump beam power.
The probe beam diameter was $d_{probe} = 0.22~\mathrm{cm}$ and its intensity $I_{probe} = 7.20~\mathrm{\mu W/cm^2}$ for all cases displayed in figure~\ref{fig:no-subtraction}.
The experimental data (blue dots) for different pump beam intensities (indicated above each subfigure of fig.~\ref{fig:no-subtraction}) were obtained by changing the pump beam power while keeping the pump diameter at a constant value of $d_{pump}=2.14~\mathrm{cm}$.
The calculated signals were obtained with a fly-through relaxation rate of $\gamma = 7~\mathrm{kHz}$ (see Eq.~\ref{eq:small_gamma}) and the Rabi frequencies $\Omega_R$ are indicated above each subfigure. 

For high pump power levels, the dispersion shape (odd function) of the measured probe absorption signals is somewhat distorted, and resembles a mixture of dispersion shape and a Lorentzian shape (even function). Nevertheless they are reasonably well described by the theoretical model (see two top subfigures of Fig.~\ref{fig:no-subtraction}).

On the other hand, for lower pump powers, the interaction of the atoms with probe beam dominates and the dispersion curve in the absorption signal is overpowered by the absorption of the linearly polarized probe beam, and so one sees mostly a Hanle resonance, which the model fails to describe (see two bottom subfigures of Fig.~\ref{fig:no-subtraction}).

\begin{figure}[htb!]
    \centering
    \includegraphics[width=0.5\linewidth]{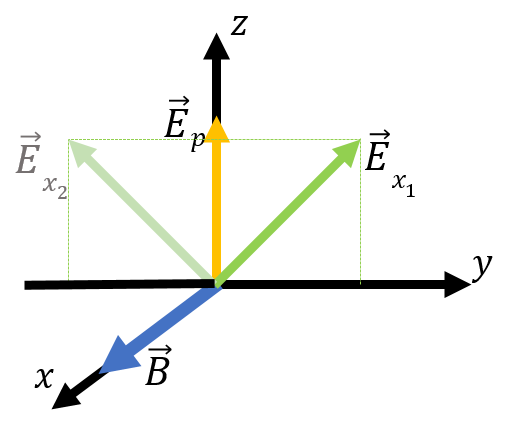}
    \caption{Pump-probe geometry for the subtraction of the even function influence on the signal. The pump and probe polarizations are all in the plane perpendicular to the magnetic field being measured. The pump beam polarization is denoted by the subscript $p$ and the two probe beam polarization orientations are denoted by $E_{x_1}$ and $E_{x_2}$}
    \label{fig:probe-subtraction-geometry}
\end{figure}
Thus the linear response (range) of the real magnetic field measurement is limited by the influence of these two factors, namely, the absolute intensities of the pump and probe beams. Furthermore the linear response is asymmetrical with respect to the zero magnetic field for higher pump intensities where the influence of the probe beam is negligible.
One solution would be to find an optimal combination of pump and probe laser intensities, but that would require precise control over both beam intensities and still the range of the linear dependence of the signal on the external magnetic field would be limited and asymmetrical.

\begin{figure}[htb!]
    \centering
    \includegraphics[width=\linewidth]{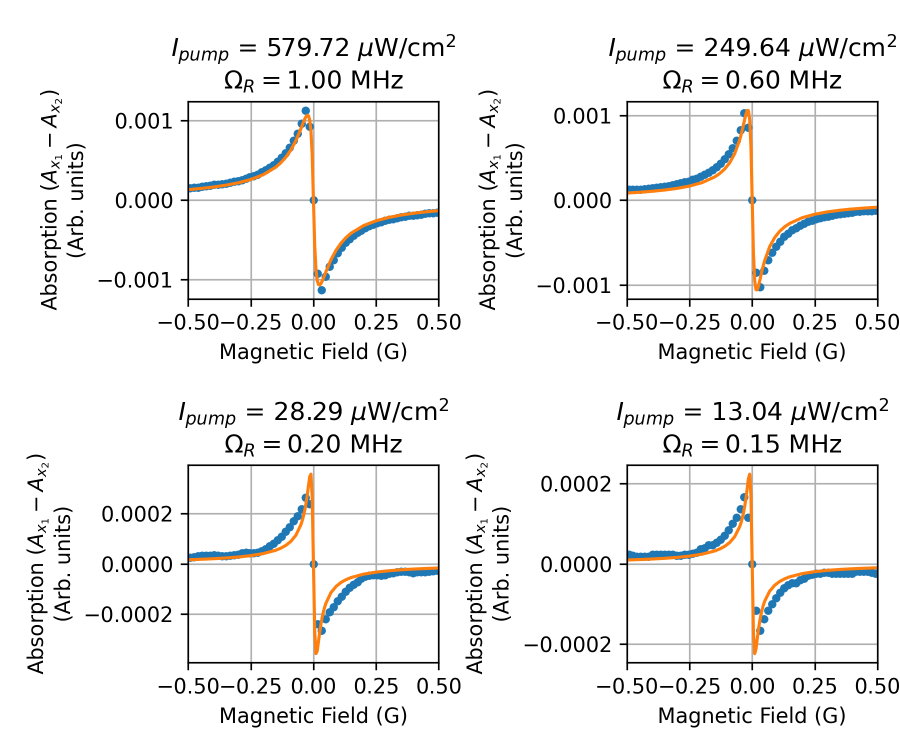}
    \caption{Probe beam absorption signals after subtracting signals for the two probe beam orientations shown in Fig.~\ref{fig:probe-subtraction-geometry} for various pump intensities for the $^{133}Cs~D_1$ $F_g = 4\rightarrow F_e = 4$ transition. Blue dots show the experimental measurements and the solid, orange line shows the results of experimental calculations.}
    \label{fig:subtracted-signals}
\end{figure}
We suggest another way to eliminate the pump and probe beam contribution to the signal that would extend the range of the linear response and be symmetrical with respect to the zero magnetic field. 
This can be done by repeating the experiment with the probe beam's polarization axis reflected around the pump beam polarization direction as shown in Fig.~\ref{fig:probe-subtraction-geometry}.
Because the two signals will produce the same signal shape except for the reversed magnetic field axis, it is not necessary to actually perform the measurement of $E_{x_2}$, it is sufficient to measure just the signal from $E_{x_1}$ and then flip the magnetic field axis to obtain the signal from $E_{x_2}$.
The two obtained probe absorption signals ($A_{x_1}$ and $A_{x_2}$) are then subtracted, one from the other ($A_{x_1}-A_{x_2}=\Delta A_{x}$).
The geometry is such that the influence of even-function part (of the pump and probe beam) is identical for both probe polarizations $E_{x_1}$ and $E_{x_2}$, and so it is canceled by the subtraction. However, the two probe beams yield opposite-sign dispersion curves (odd function part), and so the result of the subtraction doubles the amplitude.
The results can be seen in Fig.~\ref{fig:subtracted-signals}. The experimental and theoretical parameters are the same as the ones used in figure~\ref{fig:no-subtraction}, but the curves now have pure dispersion shape.
Furthermore, the calculated curves now describe the measured signals much better, also at low pump intensities.
As a result, with this treatment of the signals, the theoretical model becomes a useful tool for optimizing experimental parameters, whereas before it was limited. 

%%%%%%%%%%%%%%%%%%%%%%%%%%%%%%%%%%%%%%%%%%%%%%%%%%%%%%%%%%%%%%%%%%%%%%%%%%%%%%%%
\clearpage
\subsection{Pump power dependence} \label{sec:ppower}
\begin{figure*}[hbt!]
    \centering
    \includegraphics[width=\textwidth]{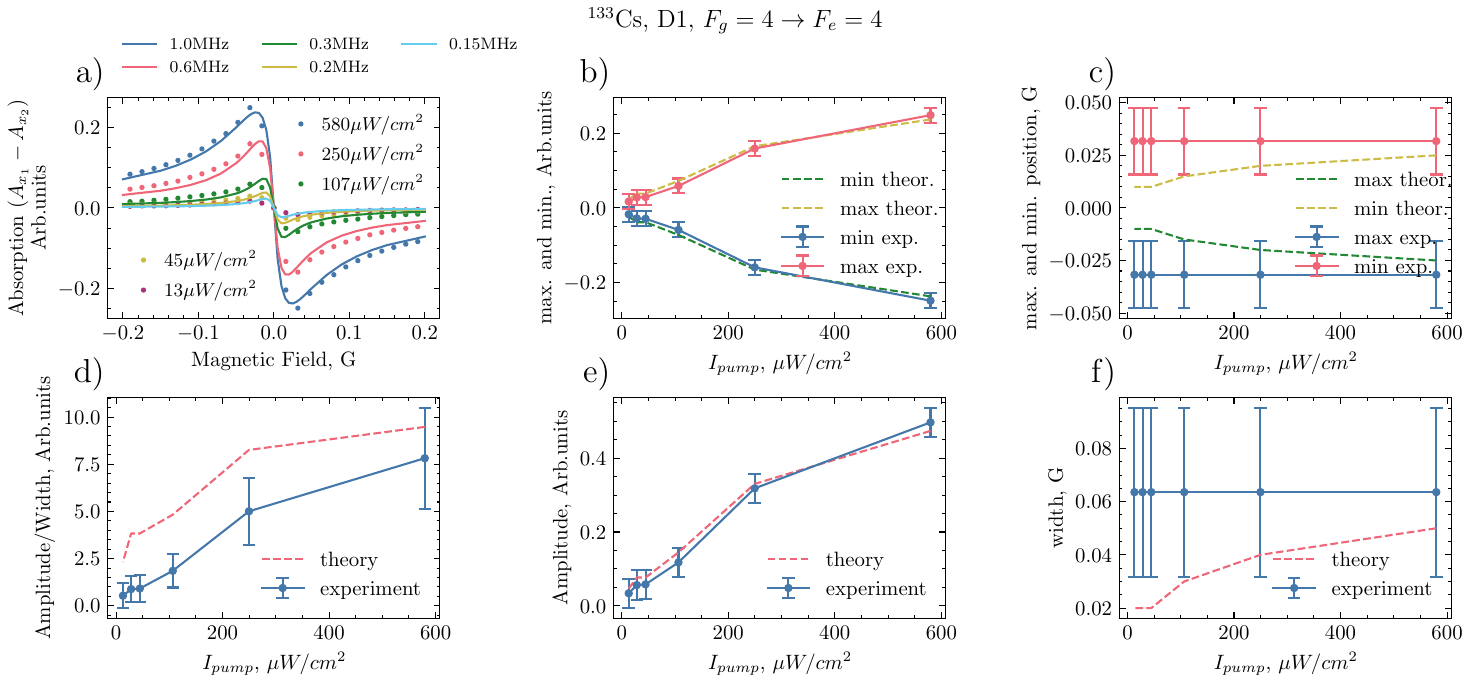}
    \caption{Influence of the pump beam intensity on the magnetometer signals on the $^{133}$Cs $D_1$ $F_g = 4 \rightarrow F_e = 4$ transition. (a) probe beam absorption ($\Delta A_x$) versus magnetic field (experimental and theoretical data). (b) Maximum and minimum values of the signal versus pump beam intensity. (c) Maximum and minimum positions with respect to the magnetic field versus pump beam intensity. (d)  Amplitude/Width ratio versus pump beam intensity. (e) Amplitude ($\Delta A_{x_{max}}-\Delta A_{x_{min}}$) versus pump beam intensity. (f) Width ($B_{\Delta A_{x_{max}}}-B_{\Delta A_{x_{min}}}$) versus pump beam intensity. In (b)--(f) the dots are the results from experimental data, while the connecting lines are only to guide the eye and the dashed lines are theoretical data.}   
    \label{fig:44_pump_pwr}
\end{figure*}

Using the signal processing method described in subsection~\ref{sec:subtraction}, it is now possible to study fruitfully the dependence of the magnetometer signals on the pump beam intensity, and the results are shown in Fig.~\ref{fig:44_pump_pwr}.
Figure~\ref{fig:44_pump_pwr}(a) shows the probe absorption signal ($\Delta A_{x}$) as a function of magnetic field for various values of the pump power.
Colored dots represent experimental measurements and solid lines show the results of the calculations.
The pump laser intensities used in the experiment are given next to data and the associated Rabi frequencies used in the calculation are given in the legend above.
For all measurements in figure~\ref{fig:44_pump_pwr} the probe beam diameter was $d_{probe}=$0.22~cm with intensity $I_{probe}=$7.2~$\mu W/cm^2$ while the pump beam diameter was $d_{pump}=$2.14 cm.
We are interested in two characteristics of this dispersion signal ($\Delta A_{x}$): (i) the amplitude, defined as the difference between the maximum and the minimum of the signal ($\Delta A_{x_{max}}-\Delta A_{x_{min}}$) and (ii) the width, defined as the difference between magnetic field values of the maximum and minimum signal values ($B_{\Delta A_{x_{max}}}-B_{\Delta A_{x_{min}}}$).

Because the signal is the difference of the two orthogonal linearly polarized absorption components in the same plane ($\Delta A_{x}=A_{x_1}-A_{x_2}$), the maxima~($\Delta A_{x_{max}}$) and minima~($\Delta A_{x_{min}}$) move away symmetrically (see Fig.~\ref{fig:44_pump_pwr}(b)) from the origin of the vertical axis in figure~\ref{fig:44_pump_pwr}(a).
The magnetic field values of the maximum ($B_{\Delta A_{x_{max}}}$) and minimum ($B_{\Delta A_{x_{min}}}$) also move away symmetrically (see Fig.~\ref{fig:44_pump_pwr}(c)) from the origin of the horizontal axis in figure~\ref{fig:44_pump_pwr}(a).
Although the experimentally determined magnetic field values of the maximum and minimum due the limited horizontal resolution show no dependence on the pump intensity in figure~\ref{fig:44_pump_pwr}(c), the experimental data are still symmetrical with respect to the origin.
This symmetrical behavior is important for achieving symmetrical operation range for positive and negative measurements of the magnetic field. The individual components show no such symmetry.

Figures~\ref{fig:44_pump_pwr}(e) and \ref{fig:44_pump_pwr}(f) show the amplitude and width of the dispersion signals, respectively.
In general, as the pump laser intensity is increased the amplitude of the signal grows (Fig.~\ref{fig:44_pump_pwr}(e)).
This can be crucial to obtaining best signal-to-noise ratio.
The theoretical data in figure~\ref{fig:44_pump_pwr}(f) shows that the width of the signal in figure~\ref{fig:44_pump_pwr}(a) grows as the optical power of the pump beam is increased, but due to low horizontal resolution this cannot be resolved in the experimental data.
The apparent offset between the experimentally and theoretically determined dispersion signal widths in figure~\ref{fig:44_pump_pwr}(f) could be attributed to some slight imperfections in determining the correct value of $\gamma$ (see Eq.~\ref{eq:small_gamma}).

Figure~\ref{fig:44_pump_pwr}(d) shows the amplitude-to-width ratio, which is a coarse estimate of the parameter that must be maximized for optimum sensitivity.
According to figure~\ref{fig:44_pump_pwr}(d), the sensitivity grows continuously with pump beam intensity, but the growth rate decreases above a pump beam intensity of about $250~\mathrm{\mu W/cm^2}$.

%%%%%%%%%%%%%%%%%%%%%%%%%%%%%%%%%%%%%%%%%%%%%%%%%%%%%%%%%%%%%%%%%%%%%%%%%%%%%%%%
\clearpage
\subsection{Pump beam diameter dependence}\label{sec:diameter}

\begin{figure*}[htb!]
    \centering
    \includegraphics[width=\textwidth]{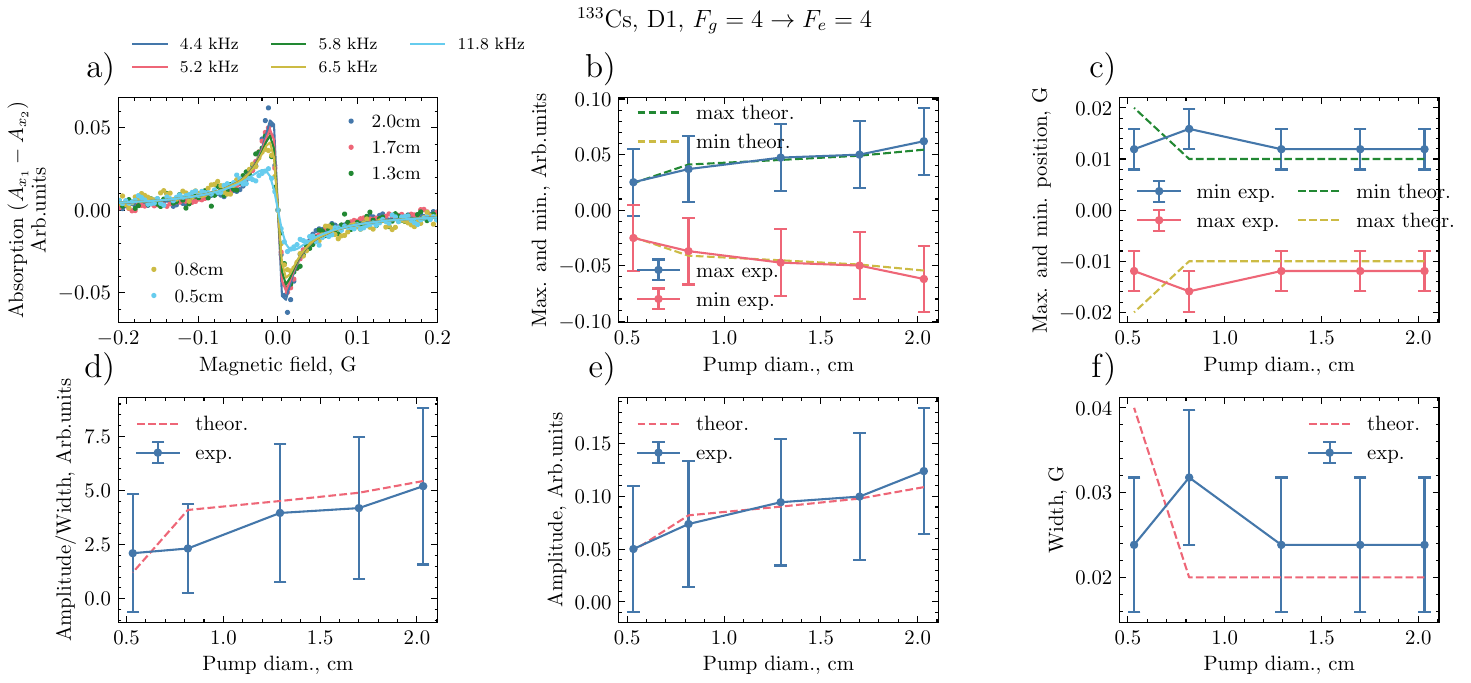}
    \caption{Probe absorption signal versus magnetic field for several values of the pump beam diameter on the $^{133}$Cs $D_1$ $F_g = 4 \rightarrow F_e = 4$ transition. (a) probe beam absorption ($\Delta A_x$) versus magnetic field (experimental and theoretical data). (b) Maximum and minimum values of the signal versus pump beam diameter. (c) Maximum and minimum positions with respect to the magnetic field versus pump beam diameter. (d)  Amplitude/Width ratio versus pump beam diameter. (e) Amplitude ($\Delta A_{x_{max}}-\Delta A_{x_{min}}$) versus pump beam diameter. (f) Width ($B_{\Delta A_{x_{max}}}-B_{\Delta A_{x_{min}}}$) versus pump beam diameter. In (b)--(f) the dots are the results from experimental data, while the connecting lines are only to guide the eye and the dashed lines are theoretical data.}    
    \label{fig:pump_diametrs}
\end{figure*}

As it was already indicated in the previous section~\ref{sec:ppower} another parameter estimation for the validation of our theoretical model is the transit relaxation rate $\gamma$.
For centimeter-sized vapor cells at room temperature this is the main cause for relaxation~\cite{Auzinsh:2013,Fabricant:2023}.
The estimate of the transit relaxation is inversely proportional to the beam diameter as was shown in Eq.~\ref{eq:small_gamma}.
Figure~\ref{fig:pump_diametrs}(a) shows the probe absorption signal ($\Delta A_{x}$) as a function of magnetic field for various values of the pump beam diameter.
Colored dots represent experimental measurements and solid lines show the results of the calculations.
The values of pump diameter ($d_{pump}$) used in the experiment are given next to data and the associated ground-state decoherence rates used in the calculation are given in the legend above.
We kept the intensity of the pump beam constant at a value of $I=\mathrm{30~\mu W/cm^2}$ and the probe beam intensity at $I=\mathrm{0.67~\mu W/cm^2}$ with $d_{probe}=\mathrm{1.95~cm}$.
All of the simulated signals where done with $\Omega_R=\mathrm{0.5~MHz}$.
Here again we are interested in two characteristics of the dispersion signal: amplitude and width, both defined the same as in section~\ref{sec:ppower}.

Figures~\ref{fig:pump_diametrs}(b) and (c) show that the positions of the maxima and minima for increasing beam diameter values also spread symmetrically with respect to the origin both along the vertical and horizontal axis of figure~\ref{fig:pump_diametrs}(a).

Figures~\ref{fig:pump_diametrs}(e) and~\ref{fig:pump_diametrs}(f) show the dependence of amplitude and width, respectively, of the dispersion signals on the pump beam diameter.
In general, as the pump beam diameter is increased the amplitude of the signal grows (Fig.~\ref{fig:pump_diametrs}(e)).
The theoretical data in figure~\ref{fig:pump_diametrs}(f) shows that the width of the signal in figure~\ref{fig:pump_diametrs}(a) decreases as the diameter of the pump beam is increased. The experimental data show similar behavior except for the case of $d_{pump}=0.5\ \mathrm{cm}$, but this could be attributed to noise affecting the precise determination of the maximum and minimum positions of this particular signal.

Figure~\ref{fig:pump_diametrs}(d) shows the amplitude-to-width ratio.
According to figure~\ref{fig:pump_diametrs}(d), the sensitivity grows continuously with pump beam diameter, but the growth rate decreases above a pump beam diameter of about $1.7~\mathrm{cm}$.

%%%%%%%%%%%%%%%%%%%%%%%%%%%%%%%%%%%%%%%%%%%%%%%%%%%%%%%%%%%%%%%%%%%%%%%%%%%%%%%%
\clearpage
\subsection{Sensitivity characterization in terms of $\gamma$, $\Omega_R$ and magnetic field direction}
\begin{figure*}
    \centering
    \includegraphics[width=\textwidth]{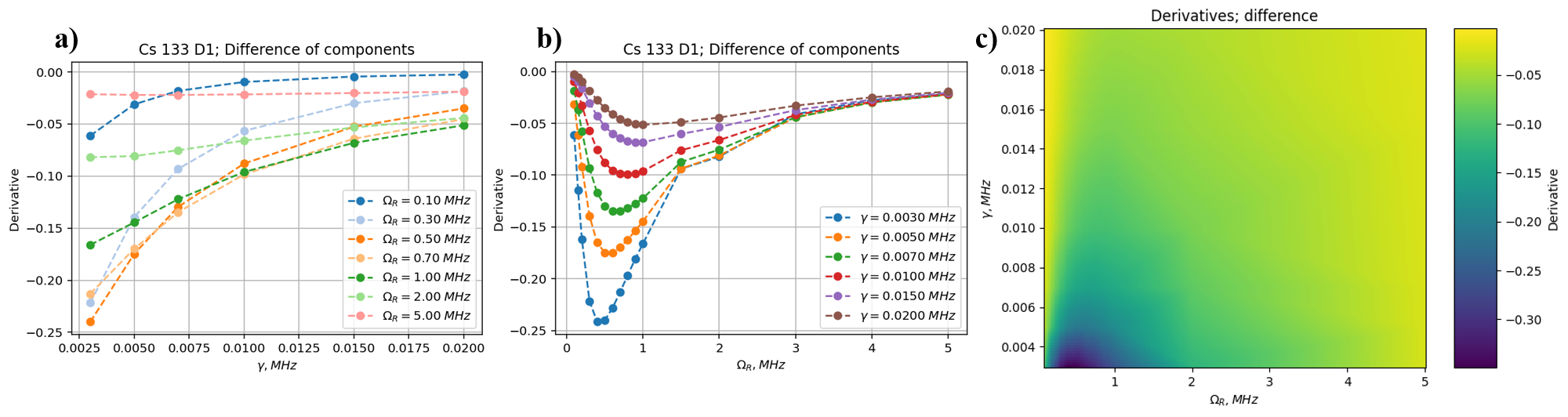}
    \caption{Relative sensitivities as derivatives at zero magnetic field dependent on a) $\gamma$; b) $\Omega_R$ and c) the dependence of the relative sensitivity on both $\gamma$ and $\Omega_R$ represented as a heat map for the $^{133}\mathrm{Cs}~\mathrm{D_1},~F_g=4\rightarrow F_e=4$ transition.}
    \label{fig:sens_param}
\end{figure*}
The results in the previous sections show good agreement between experiment and theory, which gives our theoretical model validation in describing these signals.
In this section we give the sensitivity estimates from our theoretical model.
First, we estimate the projection noise limit using
\begin{equation}
    S_x=\delta B\cdot\sqrt{t}\approx\frac{\hbar}{g_F\mu_B\sqrt{2F}}\sqrt{\frac{\gamma}{N}},
\end{equation}
from~\cite{Budker:2023}.
This gives a value of $S_x\approx1\,\mathrm{pT}/\sqrt{\mathrm{Hz}}$ for $N=2.3\cdot10^7\mathrm{}$, $\gamma=0.007~\mathrm{MHz}$ and $F=4$. This estimate serves as the lower bound of the expected measurement sensitivity and in reality could increase by about a factor of 10 when other sources of noise come in to play.

Next, we determine the impact of pump beam intensity and pump beam diameter on the relative sensitivity.
We define the relative sensitivity $s_x$ as the derivative (i.e. the slope) of the dispersion signal ($\Delta A_{x}$) at zero magnetic field.
Figure~\ref{fig:sens_param}(a) and (b) show the dependence of the signal derivative (relative sensitivity) on $\gamma$ and $\Omega_R$, respectively.
According to figure~\ref{fig:sens_param}(a) better relative sensitivities can be achieved at smaller values of $\gamma$ and as the $\gamma$ is increased worse sensitivities are expected.
While this is true for Rabi frequencies less than 1 MHz, for larger $\Omega_R$ little dependence on $\gamma$ is expected.
In figure~\ref{fig:sens_param}(b) one can observe that the absolute value of the derivative increases as the Rabi frequency is increased, but only up until it reaches the value of 0.5 MHz after which increasing the $\Omega_R$ will only diminish the relative sensitivity. Larger changes in the sensitivity can be observed for smaller $\gamma$ (blue data in fig.~\ref{fig:sens_param}(b)). It should also be noted that for larger $\Omega_R$ the choice of $\gamma$ becomes less relevant because the value of the derivatives for different $\gamma$ converge to a single value.
Thus a compromise between pump beam power ($\Omega_R$) and pump beam diameter ($\gamma$) should be determined. Figure~\ref{fig:sens_param}(c) shows a heat-map of the derivatives with respect to $\gamma$ and $\Omega_R$.
From our estimates the $\Omega_R$ should be less than 1 MHz and $\gamma$ should not exceed 0.010 MHz.
Our measurements from previous sections were in good agreement with $\Omega_R=$0.5 MHz and $\gamma=$0.007 MHz, which approximately correspond to pump beam intensity $I_{pump}=$ 200 $\mu W/cm^2$ and pump beam diameter $d_{pump}=$ 2 cm.

\begin{figure}[htb!]
    \centering
    \includegraphics[width=.5\linewidth]{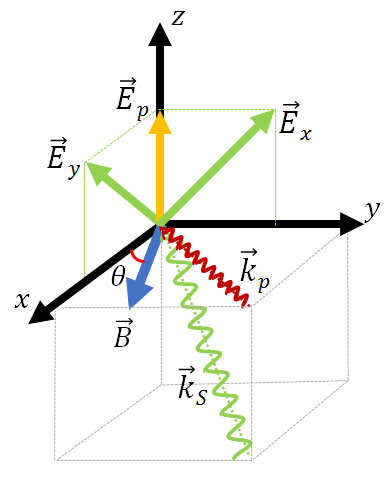}
    \caption{Geometry for 2-D magnetometry of an arbitrary magnetic field in the $xy$-plane. The pump beam polarization and propagation vectors are denoted by the subscript $p$. The probe beam propagation is denoted by the subscript $S$ and the probe polarizations are denoted by the subscript $x$ and $y$ as in Fig.~\ref{fig:geometry}. The angle between the external magnetic field and the $x$-axis is denoted by $\theta$.}    
    \label{fig:geometry_theta}
\end{figure}

\begin{figure}
    \centering
    \includegraphics[width=\linewidth]{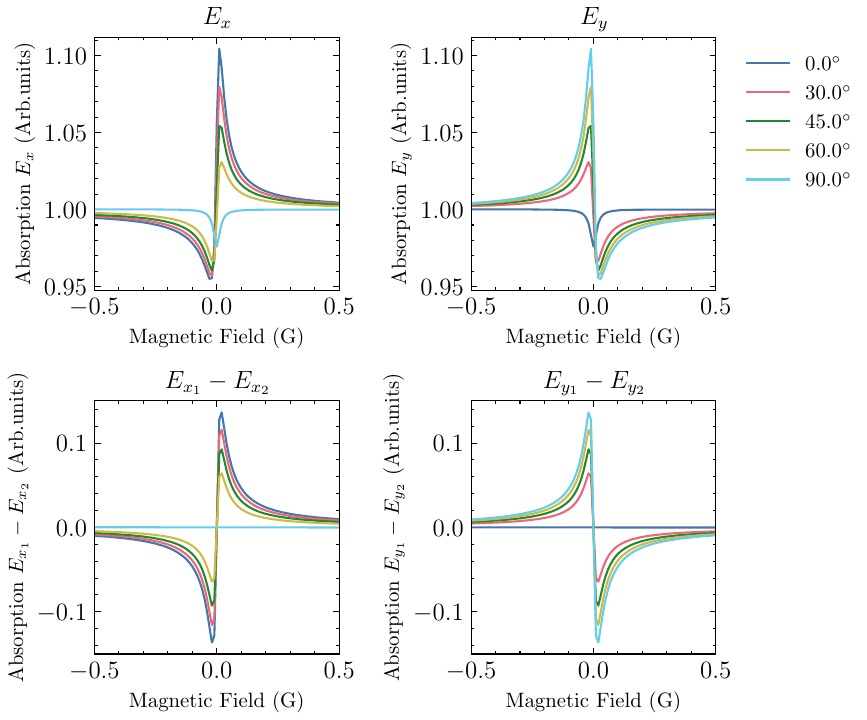}
    \caption{Numerically calculated probe beam absorption signals for $\vec{E}_x$ and $\vec{E}_y$ probe orientations (top) and differences of symmetrical probe components (bottom) as a function of magnetic field for various angles $\theta$ for the $^{133}\mathrm{Cs}~\mathrm{D_1},~F_g=4\rightarrow F_e=4$ transition.}    
    \label{fig:B_rotation}
\end{figure}

By using an EOM to alternate between $\vec{E}_x$ and $\vec{E}_y$ orientations of the probe beam polarization (see Fig.~\ref{fig:geometry_theta}), one can determine the magnitude and direction of an arbitrary magnetic field in the $xy$-plane.
Fig.~\ref{fig:B_rotation} shows the results of numerical simulations of the probe absorption versus magnetic field magnitude for different angles $\theta$ between the magnetic field vector and the $x$-axis for the $^{133}\mathrm{Cs}~\mathrm{D_1},~F_g=4\rightarrow F_e=4$ transition.
The signals were modeled for $\Omega_R=~0.5$ and $\gamma=7~\mathrm{kHz}$.
The top left plot shows the results for the $\vec{E}_x$ polarization of the probe laser at different angles, whereas the right top plot shows results for $\vec{E}_y$ polarization.

In general, the absorption signal of an individual probe polarization component takes the form of an asymmetrical dispersion signal with respect to the vertical axis, except in the case of $\theta=45^{\circ}$, in which case the signal is completely symmetrical with respect to the vertical axis and another special case of $\theta=0^{\circ}$, in which the dispersion shape has been transformed into a very narrow peak centered around $B=0$.
This feature gives a useful criterion for aligning the magnetic field along a given axis and determining when it is precisely zero.

\begin{figure}
    \centering
    \includegraphics[width=\linewidth]{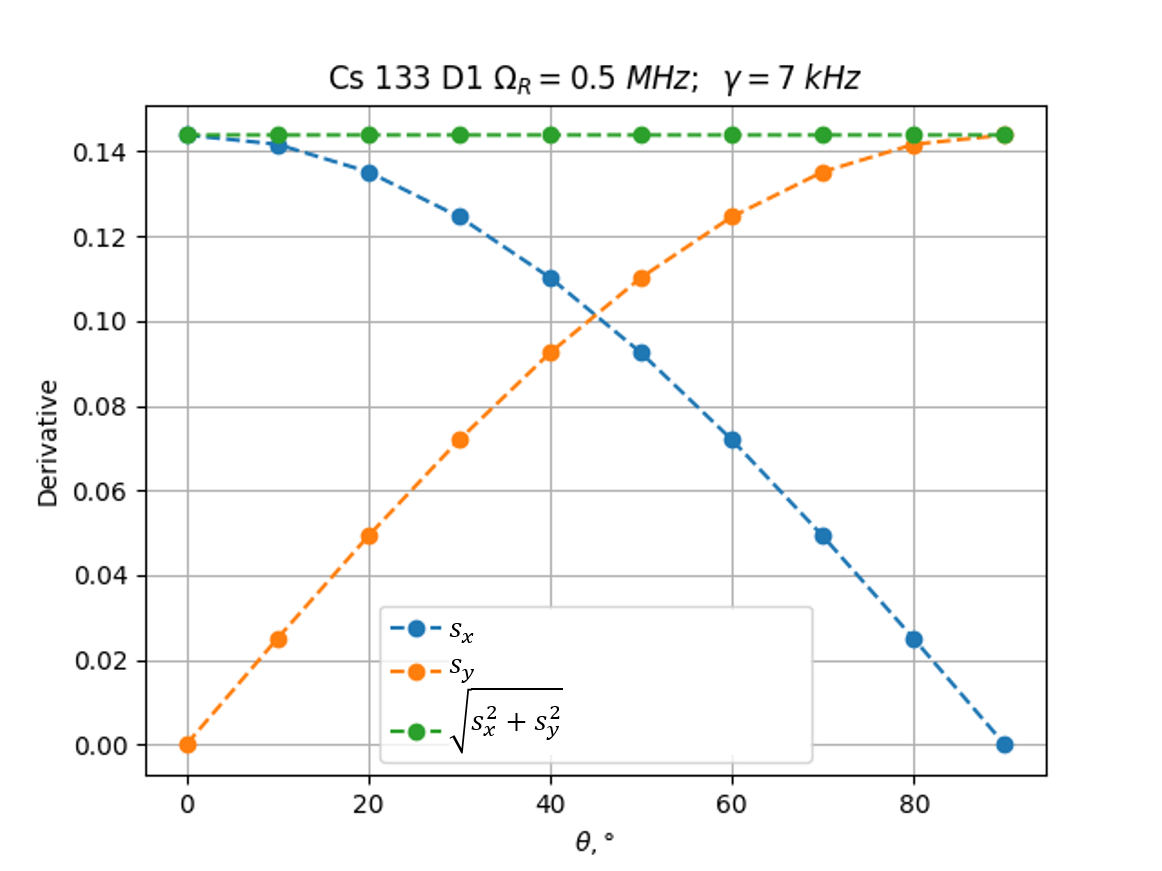}
    \caption{Relative sensitivities as derivatives at zero magnetic field dependent on the angle $\theta$ for the $^{133}\mathrm{Cs}~\mathrm{D_1},~F_g=4\rightarrow F_e=4$ transition.}
    \label{fig:sens_theta}
\end{figure}

The bottom two plots in figure~\ref{fig:B_rotation} show the absorption signal as the difference between the two orthogonal components in the $zy$-plane (bottom left) and $zx$-plane (bottom right) as explained in \ref{sec:subtraction}.
In this case the dependence on the magnetic field angle $\theta$ leads to a decrease in the signal amplitude as $\theta$ is increased for $\Delta A_x$ and the reverse response for $\Delta A_y$.
This means that as the relative sensitivity decreases for one probe polarization component, it increase for the other.
The dependence of the relative sensitivities on the magnetic field angle $\theta$ is shown in figure~\ref{fig:sens_theta}.
Because the proposed setup (see sec.~\ref{sec:experiment}) allows for simultaneous detection of both probe polarization components ($\vec{E}_x$ and $\vec{E}_y$) the total sensitivity stays constant for all values of $\theta$ (see fig.~\ref{fig:sens_theta} green data) when determining the magnitude and angle of the magnetic field in the $xy$-plane.

%%%%%%%%%%%%%%%%%%%%%%%%%%%%%%%%%%%%%%%%%%%%%%%%%%%%%%%%%%%%%%%%%%%%%%%%%%%%%%%%
\clearpage
\section{Conclusions}
We have adapted the two-axis Hanle magnetometer proposed by LeGal~\cite{LeGal:2019} for Helium to the Cesium $D_1$ line.
By measuring probe beam absorption signals with two orthogonal probe beam polarizations and subtracting these signals, it was possible to eliminate systematic background caused by remnant Hanle absorption signal of the probe beam.
As a result, the measured experimental signals agreed well with signals that were calculated based on a theoretical model.
The strength of the dispersion signal was compared for the different transitions.
The dependence of the amplitude, width, and amplitude-to-width ratio of the dispersion signals on pump power and pump beam diameter were studied for the $F_g=4\longrightarrow F_e=4$ transition.
We determined the optimum values for pump-beam power and pump-beam diameter.
In principle, the projection-noise limit for this transition could be estimated to be on the order of $\approx 1\,\mathrm{pT}/\sqrt{\mathrm{Hz}}$.
These values are already competitive with many commercial magnetometers and could be used for applications such as detecting magnetic anomalies, navigating in the earth's magnetic field, or measuring biological magnetic signals.
The sensitivities obtained in this work could be further improved if a coated cell or buffer gas is used. Heating the cell could also prove beneficial.

\begin{acknowledgments}
We acknowledge support from Latvian Council of Science project “Compact 3-D magnetometry in Cs atomic vapor at room temperature” Project No. lzp-2020/1-0180.
\end{acknowledgments}

\bibliography{Hanle-2024}

\end{document}